\documentstyle[12pt]{article}
\newcommand{\beq}[1]{\begin{equation}\label{#1}}
\newcommand{\eeq}   {\end{equation}}

\begin{document}
\begin{center}

{\bf
MULTIPLE SCATTERING ERROR PROPAGATION IN PARTICLE TRACK RECONSTRUCTION
}

\bigskip\bigskip

\normalsize M.~Pen\c{t}ia, G.~Iorgovan \\
\medskip
INSTITUTE OF ATOMIC PHYSICS \\
RO-76900 P.O.Box MG-6, Bucharest, ROMANIA \\
e-mail: pentia@roifa.bitnet \\
\vspace{1cm}
\normalsize A.~Mihul \\
\medskip
BUCHAREST UNIVERSITY \\
RO-76900 P.O.Box MG-11, Bucharest, ROMANIA \\
e-mail: mihul@roifa.bitnet \\

\vspace{5cm}
\begin{abstract}
       Particle track  reconstruction  capabilities
       of the silicon  tracking  detector  system  have  been
       studied. As  the  multiple  Coulomb  scattering  (MCS)
       induces unavoidable uncertainties  on  the  coordinate
       measurement, the corresponding error estimates and the
       associated correlations have been  used  to  find  the
       best track fit parameters and their errors. Finally it
       permits to find the proper  particle  characteristics,
       as  vertex position  and  resolution, flight direction
       and the error.
\end{abstract}

\end{center}

\vfill
\clearpage
\pagestyle{myheadings}
\setcounter{page}{1}
\section {Introduction}

Design and preparation of any components of a  detector  system
must  take  care  of  characteristics  and  detection   performances
(efficiency, acceptance, position or  energy  resolution)  necessary
for a specific process study. The silicon tracking detector  system,
in our case, must furnish the best information about the  coordinate
track intercept of the incident particle on  every  detector  layer.
Otherwise, for a good track reconstruction, it is necessary to  find
the best estimate of the track parameters  in  a  specific  particle
measurement. This is the reason we studied, the  particle  transport
in a silicon tracking system  and  estimated  the  multiple  Coulomb
scattering (MCS) perturbation in particle track measurements.

This study furnishes information on the position resolution  we
could get from every detector layer, and also its dependence on some
physical and geometrical parameters. In the detector  system  design
these parameters could be choosed in an optimal manner,  aiming  the
best track reconstruction possibilities or the best estimate of  the
track parameters (vertex position, curve radius) directly  connected
to physical quantities.

The simplified tracking system we used, consists of a 5 layers
silicon microstrip detector, $300 \mu m$ thick, interspaced by 1.5~cm,
and the first detector layer located at 130~cm from the  interaction
point (see Fig.1).

\section{The track particle position uncertainties due to
                    multiple Coulomb scattering.}

When a charged particle is traversing the detector elements  of
a tracking system, it undergoes small deviations of the  track,  due
to MCS. The effect is usually described by  the  theory  of  Moliere
(see for example \cite{1}). It shows that, by traversing  detector's
material, thickness $s$, the particle undergoes successive small-angle
deflections, symmetrically distributed about the incident direction.
Applying the central limit theorem of statistics to a  large  number
of independent scattering events, the Moliere  distribution  of  the
scattering angle can be approximated by a Gaussian one \cite{2}.  It
is sufficient for many applications to use Gaussian approximation for
the central 98\% of the  plan  projected  angular  distribution.  The
width of this distribution is the root mean square of the scattering
angle \cite{3}
\beq{eq1}
\theta_0 = \frac{13.6MeV}{p \beta c} z_c \sqrt{\frac{s}{X_L}}
\left[1 + 0.038~ln\left(\frac{s}{X_L}\right)\right]
\eeq

\noindent where $p, \beta c$ and $z_c$ are the momentum, velocity and charge
number of the incident particle,  and  $X_L$ is the radiation length of the
scattering medium. That is, the plane projected angle $\theta_{plane,x}$ or
$\theta_{plane,y}$ of the deflection angle $\theta$, onto the xOz and yOz
planes, where the x and y axes are orthogonal to the Oz direction of
motion, shows an approximately Gaussian angular distribution
\beq{eq2}
\frac{1}{\sqrt{2\pi}\theta_0}exp\left[-\frac{\theta_{plane}^2}{2\theta_0^2}
\right]d\theta_{plane}
\eeq
Deflections  into  $\theta_{plane,x}$ and $\theta_{plane,y}$
are independent and identically distributed, and
$\theta_{space}^2 = \theta_{plane,x}^2 + \theta_{plane,y}^2$.

The  angular  distribution  is  translated  to   a   coordinate
distribution by particles fly onto every detector  layer.  The  more
intersected  layers the  larger  distribution  width is. The
coordinate distribution is defined by statistical spread due to MCS,
and depends on the number and position of the  intersected  detector
layer elements. It has the same form as angular distribution

\beq{eq3}
\frac{1}{\sqrt{2\pi}\sigma_{x_i}}exp\left[-\frac{x_i^2}{2\sigma_{x_i}^2}
\right]~dx_i
\eeq

\noindent with the mean square deviation (distribution width) as the
squares sum of the $(i-1)$ preceding distribution widths projected onto
$i$-th detector layer
\beq{eq4}
\sigma_{x_i}^2 \equiv <x_i^2> =
\theta_0^2\left[(z_i-z_1)^2 + (z_i-z_2)^2+...+(z_i-z_{i-1})^2\right]
\eeq
For an oblique incidence ($\theta \not= 0$) the effective path
length  in the silicon detector  is  larger,  and  the  same  is  the
position distribution width on the next  detector  layers.  Nevertheless,
in this work we will  consider  only  the  minimal  width (\ref{eq4}), as
to emphasize the precision limit in particle position measurement with
a given silicon tracking system.

\section{The Monte-Carlo particle scattering description.}

The change of track parameters is usually \cite{4} parametrized by
two mutually orthogonal, uncorrelated scattering angles
$\delta \theta _{plane,x}$ and $\delta \theta_{plane,y}$ or
($\delta \theta_{space}$  and $\delta \varphi$) which leads to a
corresponding displacement $\delta x_{plane}$ and $\delta y_{plane}$
in the position.

Following the  stochastic  nature  of  the  MCS,  we  use  the
Monte-Carlo study  by  generating  the  joint  ($\delta x_{plane},
\delta \theta _{plane,x}$) distribution with independent Gaussian
random variables ($w_1 ,w_2$) \cite{4}
\begin{eqnarray}\label{5}
\delta x_{plane}&=&\frac{w_1 s \theta_0}{\sqrt{12}}+
\frac{w_2 s \theta_0}{2} \nonumber\\
\delta \theta_{plane,x}&=&w_2 \theta_0
\end{eqnarray}

The same have been used for the joint
($\delta y_{plane},\delta \theta _{plane,y})$
distribution in yOz plane.  Finally  we  constructed  the  incidence
points distribution (coordinate distribution) on the detector  layer
no. 2, 3, 4 and 5. (see Fig.2 and Table~\ref{tbl1}), for an
incident 500~MeV/c muon.

In  Monte-Carlo  simulated  particle  transport  the   particle
position uncertainty on every detector layer has  been  measured  as
the variance or the mean square deviation  of  the  scattered  track
incidence points $x_m$ about the unscattered one $x_0$
\beq{eq6}
\sigma_x^2 = \sum \limits_{n=1}^{N} \frac{(x_n-x_0)^2}{N}
\eeq

and similar for $\sigma_y^2$ variance, for every detector layer. The
$x_n$ (and $y_n$) are the coordinates of the scattered track incidence
points on detector layer, and N is the total number of generated events.

The distribution widths $\sigma_{x_i}$ of the muon track $x$-coordinate
points on detector layer $i$= 2, 3, 4, 5 and for momentum values from 500
to 7500~MeV/c, are presented in Table~\ref{tbl1}. There are also the
analytical estimation of the same widths of the MCS incidence point
distribution on each of the $i$-detector layer, according (\ref{eq4})
\beq{eq7}
\sigma_{x_i}^2 = \frac{i(i-1)(2i-1)}{6} (l\theta_0)^2
\eeq

    where  $l$ - the distance between detector layers ($l=1.5$ cm)
     and   $\theta_0$ - the plane r.m.s. scattering angle (\ref{eq1}).

This Monte-Carlo particle transport description within silicon detector
system will be used as a coordinate data generator for the track
reconstruction procedure.

\section{Position error correlations.}

MCS produces errors correlated from  one  layer  to  the  next.
Clearly a scattering in layer~1 produces correlated position  errors
in layer~2, 3 and so on (see Fig.3).  The  proper  error  matrix  is
non-diagonal \cite{5}, and it must be find out.

Let's denote $\delta x_i$ the track deviation $x$-coordinate point on
the $i$-th layer, with respect to the initial incident direction on the
detector system, then (see Fig.3)
\beq{eq8}
\begin{array}{lll}
&\delta x_1=0& \\
&\delta x_2=\delta _1& \\
&\delta x_3=\delta_{31}+\delta_2& \\
&\delta x_4=\delta_{41}+\delta_{42}+\delta_3& \\
&\delta x_5=\delta_{51}+\delta_{52}+\delta_{53}+\delta_4&
\end{array}
\eeq
where the individual contributions due to preceding scatterings are
\begin{eqnarray}\label{eq9}
\delta_{31}&=&\delta_1 \frac{z_3-z_1}{z_2-z_1} \nonumber\\
\delta_{41}&=&\delta_1 \frac{z_4-z_1}{z_2-z_1}~~;~~
\delta_{42}=\delta_2 \frac{z_4-z_2}{z_3-z_2} \\
\delta_{51}&=&\delta_1 \frac{z_5-z_1}{z_2-z_1}~~;~~
\delta_{52}=\delta_2 \frac{z_5-z_2}{z_3-z_2}~~;~~
\delta_{53}=\delta_3 \frac{z_5-z_3}{z_4-z_3} \nonumber
\end{eqnarray}
Because they are statistical variables it is necessary to
find their mean value, and to express it by the independent scattering
deviations $\delta_k$, unaffected by earlier scatterings
\beq{eq10}
\delta_k \equiv \sqrt{<\delta_k^2>} = \theta_0 (z_{k+1}-z_k)
\eeq

We try to express the $\delta x_i$ deviations by the independent
$\delta_k$  ones (\ref{eq10}), for which $<\delta_k>=0$
\beq{eq11}
\delta{x_i}=\sum \limits_{k=1}^{i-1} \delta_k\frac{z_i-z_k}{z_{k+1}-z_k}
\eeq

Now the position error (covariance) matrix ($V_{ij}$),
defined \cite{4} as the statistical mean of the pair deviation products
$<\delta x_i \delta x_j>$ for all possible detection layers is
\beq{eq12}
\begin{array}{ll}
{}~~V_{ij}&\equiv <\delta x_i \delta x_j> = \\
&=\theta_0^2\left[ (z_i-z_1)(z_j-z_1)+(z_i-z_2)(z_j-z_2)+...+
(z_i-z_{i-1})(z_j-z_{i-1}) \right] \\ \medskip
for&$i$\leq $j$ = 1,2,...,$n$ \\
\end{array}
\eeq
The $(V_{ij})$ matrix is symmetric. The error correlation matrix is
immediately
\beq{eq13}
\rho_{ij}=\frac{<\delta x_i\delta x_j>}{\sqrt{<\delta x_i^2><\delta x_j^2>}}
\eeq
The uncorrelated position errors in the coordinate reading $\sigma_0$,
have to be added in squares ($\sigma_0^2$) into the diagonal terms of
the error matrix V.

For example, for a 500~MeV/c MUON track detection ($\theta_0=1.23 mrad$)
by our system configuration ($z_1 =130 cm, z_2 =131.5 cm, z_3 =133 cm,
z_4 =134.5 cm, z_5 =136 cm$ and $\sigma_0 =10 \mu m$) the matrix elements
$V_{ij}$ (in square microns) and $\rho_{ij}$ are
\beq{eq14}
V~=~
\left(
\begin{array}{ccccc}
100.&   0. & 0.  &  0.  &  0.  \\
0.  & 440.64  &  681.27  &  1021.9  &  1362.5 \\
0.  & 681.27  &  1803.2  &  2725.1  &  3747.0 \\
0.  & 1021.9  &  2725.1  &  4868.9  &  6812.7 \\
0.  & 1362.5  &  3747.0  &  6812.7  &  10319.
\end{array}
\right)
\eeq
\beq{eq15}
\rho~=~
\left(
\begin{array}{ccccc}
1 & 0 & 0 & 0 & 0 \nonumber\\
0 & 1 & .76430 & .69768 & .63898 \\
0 & .76430 & 1 & .91970 & .86865 \\
0 & .69768 & .91970 & 1 & .96114 \\
0 & .63898 & .86865 & .96114 & 1
\end{array}
\right)
\eeq
\noindent As long as the $V_{ij}$  matrix elements depend both  on
kinematical characteristics of  the  detected  particles and on the
tracking detector system configuration, the $\rho_{ij}$ matrix elements
are independent on particle characteristics, and is defined only by
system configuration.

In the following we  will  use  these  matrices  in  the  track
reconstruction by a least squares fit procedure.

\section{Track reconstruction parameters and the errors}

In the absence of the magnetic field, the unscattered track is
a straight line. The independent description of $x$ and $y$ MCS  data
permits a separate least squares fit  to  these  data  by  a  linear
relationship \cite{6,7}
\beq{eq16}
\begin{array}{lll}
x & = & x_0 + \alpha_x z \\
y & = & y_0 + \alpha_y z
\end{array}
\eeq

\noindent With the coordinate and error data $(x_i \pm \sigma_{x_i}),
(y_i \pm \sigma_{y_i}), z_i$, along with the corresponding correlation
matrix $\rho_{ij}~~(V_{ij} = \rho_{ij} \sigma_{x_i} \sigma_{x_j})$ as
input data, it is possible to express the $\chi^2$ in matrix form,
for every coordinate data set. For $x$-data set it will be
\beq{eq17}
\chi^2 = (X-HA_x)^TV^{-1}(X-HA_x)
\eeq
where
\newcommand{\colfour}[4]{\begin{array}{c} $$#1$$ \\$$#2$$ \\$$#3$$ \\$$#4$$
\end
{array}}
\newcommand{\coltwo}[2]{\begin{array}{c} $$#1$$ \\$$#2$$\end{array}}
\beq{eq18}
X=\left(\colfour{x_1}{x_2}{\vdots}{x_n} \right)~~;~~
H=\left(\colfour{1}{1}{\vdots}{1}
        \colfour{z_1}{z_2}{\vdots}{z_n} \right)~~;~~
A_x=\left(\coltwo{x_0}{\alpha_x} \right)
\eeq
\noindent the V matrix is given by (\ref{eq12}) and (\ref{eq14}).

Least squares criterion impose
\begin{displaymath}
\frac{\partial \chi^2}{\partial A_x} = 0 \mbox {~~~~~~~or~~~~~~~}
H^T V^{-1} (X - H A_x) = 0
\end{displaymath}

By solving the linear system relative to $A_x$  we get the fit parameters
\beq{eq19}
A_x = (H^T V^{-1} H)^{-1} (H^T V^{-1} X)
\eeq
and the errors of these parameters
\beq{eq20}
E_{Ax} \equiv <\delta A_x \delta A_x^T> = (H^T V^{-1} H)^{-1}
\eeq
\noindent The same procedure we applied to $y$ coordinate, as to find
the fit parameters and errors; $z_i$ has no any uncertainty,
they are detector position coordinates.

Let's take an example. Using the particle transport simulation
data within silicon tracking system, along with the variance matrix data
$V$ (\ref{eq14}), we have been obtained the fit parameters and their errors
($x_0 \pm \sigma_{x_0}, \alpha_x \pm \sigma_{\alpha_x};
  y_0 \pm \sigma_{y_0}, \alpha_y \pm \sigma_{\alpha_y}$).
The results are presented in Table~\ref{tbl2}.

The found track fit parameters, could be connected with
some particle characteristic quantities as
$xy$-vertex position ($x_0, y_0$) and the $xy$-vertex uncertainty
($\sigma_{xy}^{(vertex)} = \sqrt{\sigma_{x_0}^2 + \sigma_{y_0}^2}$)
and the particle production direction
($\theta \pm \sigma_{\theta}, \varphi \pm \sigma_{\varphi}$). This
last one is connected with the slope parameters $\alpha_x$ and $\alpha_y$.
See Table \ref{tbl2}.
\begin{eqnarray}\label{eq21}
tg\varphi&=&{\alpha_y\over \alpha_x} \nonumber\\
tg\theta&=&{\alpha_x\over cos\varphi}
\end{eqnarray}
\noindent and the corresponding errors
\begin{eqnarray}\label{eq22}
\sigma_{tg\varphi}^2&=&\Biggl({1\over\alpha_x}\Biggr)^2 \sigma_{\alpha_y}^2 +
\Biggl({\alpha_y\over\alpha_x^2}\Biggr)^2 \sigma_{\alpha_x}^2 \nonumber\\
\sigma_{tg\theta}^2&=&\Biggl({1\over cos\varphi}\Biggr)^2 \sigma_{\alpha_x}^2 +
\Biggl({\alpha_x\over cos^2\varphi}\Biggr)^2 \sigma_{cos\varphi}^2
\end{eqnarray}
In Table~\ref{tbl3} there  are  results  on  the reconstructed $\theta $ and
$\varphi $ direction values in comparison with the generated ones.

The $xy$-vertex position uncertainty, due to MCS in the detector material,
have been also  calculated as
$\sigma_{xy}^{(vertex)}=\sqrt{\sigma_{x_0}^2+\sigma_{y_0}^2}$
and  depends, of  course,  on  particle
momentum. If we want to reduce this error it is necessary  to  bring
closer the detector tracking system relative to  interaction  point.
The $xy$-vertex position uncertainty (resolution) dependence on particle
(muon) momentum and $z_1$ distance to first detector layer is shown
in Fig.4.
The result is useful in detector system design to  find  an  optimal
configuration in preparing the experimental work. Also,  to  have  a
choice for vertex  resolution  as  a  compromise  between  the  best
possible values in the proximity of the interaction  point  and  the
worse ones far from this point at the radiation harmless distance.
The $xy$-vertex  position  resolution due  to  MCS  errors  combined
with  the  intrinsic
detector coordinate uncertainties, could  not  be  better  than  the
values shown in Table~\ref{tbl4}. Nevertheless, from Table~\ref{tbl4}
and Fig.4 we see
also that it is possible to have a better vertex position resolution
if the detector tracking system is placed to a smaller distance from
the interaction point.

\section{Conclusions}

We have calculated the uncertainties in particle characteristic parameters
as vertex position and the particle flight direction, by a linear fitting
procedure, using position error correlation due to MCS in succesive
detector layers of a silicon tracking system.

The most important result shows these uncertainties are independent on
individual track coordinates (see for example the errors for $x_0$,$y_0$,
and $\alpha_x$,$\alpha_y$) and they depend exclusively (see (\ref{eq20})
and (\ref{eq22})) on the detector system configuration (geometry).
The results are useful both in particle tracking system design and in
experimental data analysis for pattern track recognition. The former
one could use our analytical relations for particle parameter
error estimation relative to diferent tracking system configuration.
The later one could use our data fitting procedure, including system
configuration dependence on the coordinate covariance matrix, in particle
track reconstruction and the associated parameter determination.

\section*{Acknowledgements}

One of us (MP) would like to thank Professor George Trilling (LBL) for
helpful comments on some tracking aspects.
This work was performed at the Institute of Atomic Physics, under
contract no.77B/A-53.
\vfill

\vfill
\begin{table}
\caption{       \label{tbl1}
MUON x-distribution width on the detector layers of a silicon tracking
system due to multiple Coulomb scattering.
}
\begin{center}
These are the correlated uncertainties $\sigma_i$, used in track
fitting procedure.\\
\vspace{.5cm}
MONTE-CARLO SIMULATED DATA / ANALYTICAL CALCULATED DATA (\ref{eq7})\\
\bigskip
\begin{tabular}{||c||c|c|c|c||} \hline \hline
$p$    &  $\sigma_2$ & $\sigma_3$ & $\sigma_4$ & $\sigma_5$ \\
$(MeV/c)$ & ($\mu m$) & ($\mu m$) & ($\mu m$) & ($\mu m$) \\ \hline
 500 & 18.88 / 18.46 & 41.75 / 41.27 & 69.74 / 69.06 & 101.94 / 101.09 \\
1000 &  9.20 / 9.08 & 20.60 / 20.30 & 34.43 / 33.97 &  50.34 / 49.73 \\
1500 &  6.10 / 6.03 & 13.55 / 13.49 & 22.73 / 22.58 & 33.42 / 33.05 \\
2000 &  4.56 / 4.52 & 10.19 / 10.11 & 17.05 / 16.91 & 25.07 / 24.76 \\
2500 &  3.71 / 3.61 &  8.18 / 8.08 & 13.65 / 13.53 & 19.96 / 19.80 \\
3000 & 3.08 / 3.01 & 6.84 / 6.73 & 11.36 / 11.27 & 16.57 / 16.49 \\
3500 & 2.62 / 2.58 & 5.84 / 5.77 & 9.72 / 9.66 & 14.18 / 14.14 \\
4000 & 2.28 / 2.26 & 5.09 / 5.05 & 8.54 / 8.45 & 12.52 / 12.36 \\
4500 & 2.03 / 2.01 & 4.50 / 4.49 & 7.48 / 7.51 & 10.90 / 10.99 \\
5000 & 1.82 / 1.81 & 4.05 / 4.04 & 6.79 / 6.76 & 9.96 / 9.89 \\
5500 & 1.66 / 1.64 & 3.68 / 3.67 & 6.12 / 6.14 & 8.94 / 8.99 \\
6000 & 1.53 / 1.51 & 3.39 / 3.37 & 5.67 / 5.63 & 8.31 / 8.24 \\
6500 & 1.39 / 1.39 & 3.12 / 3.11 & 5.22 / 5.20 & 7.61 / 7.61 \\
7000 & 1.31 / 1.29 & 2.90 / 2.88 & 4.85 / 4.83 & 7.06 / 7.07 \\
7500 & 1.21 / 1.20 & 2.69 / 2.69 & 4.50 / 4.50 & 6.58 / 6.59 \\ \hline \hline
\end{tabular}

\end{center}
\end{table}
\begin{table}
\caption{ \label{tbl2}
Particle transport simulation data within silicon tracking system.
}
\begin{center}
    MUON $500~MeV/c, x_0 = 0, y_0 = 0, z_0 = 0, \theta = 15^\circ,
    \varphi = 30^\circ $\\
    Silicon layer width = $300 \mu m, z_1 = 130 cm, dz = 1.5 cm $\\
    Mean scattering angle (plan projected) $\theta_0 = 1.23 mrad$\\
\bigskip
\begin{tabular}{|c|ccc|ccc|c|} \hline
  det & $x(cm)$ & $\sigma_0 (\mu m)$ & $\sigma_{MCS} (\mu m)$
      & $y(cm)$ & $\sigma_0 (\mu m)$ & $\sigma_{MCS} (\mu m)$
      & $z(cm)$\\ \hline
1 & 30.166605 & 10 &  0.0 & 17.416698 & 10 &  0.0 & 130.0 \\
2 & 30.517353 & 10 & 18.5 & 17.617701 & 10 & 18.5 & 131.5 \\
3 & 30.868230 & 10 & 41.3 & 17.816364 & 10 & 41.3 & 133.0 \\
4 & 31.218312 & 10 & 69.1 & 18.013349 & 10 & 69.1 & 134.5 \\
5 & 31.569294 & 10 & 101.1 & 18.207651 & 10 & 101.1 & 136.0 \\ \hline
\end{tabular}
\bigskip

{\bf Linear track reconstructed (parameter) data} \\
\medskip
\begin{tabular}{ll}
$x_0$=$-2316.4 \pm 1828.3 \mu m$ & $\alpha_x$=$0.23383 \pm 0.001404$\\
$y_0$=$~~618.69 \pm 1828.3 \mu m$ & $\alpha_y$=$0.13350 \pm 0.001404$\\
\end{tabular}
\medskip \\
$\sigma^{(vertex)}_{xy} = \sqrt2 \sigma_{x_0} = 2585.65 \mu m$ \\
\medskip
$\theta=15.070 ^\circ \pm 0.096 ^\circ$ \\
$\varphi=29.723 ^\circ \pm 0.298 ^\circ$ \\
\end{center}
\end{table}
\begin{table}
\caption{       \label{tbl3}
Track reconstructed direction by least squares fit of the 500 MeV/c MUON
Monte-Carlo simulated transport data within silicon tracking sistem.
}
\begin{center}
\bigskip
\begin{tabular}{||c|c||c|c||} \hline \hline
\multicolumn{2}{||c||}{Generated} & \multicolumn{2}{c||}{Reconstructed} \\
\hline
$\theta (deg)$ & $\varphi (deg)$ & $\theta (deg)$ & $\varphi (deg)$ \\ \hline
5 & 270 & $5.07 \pm 0.81$ & $269.20 \pm 0.91$ \\
10 & 240 & $10.16 \pm 0.20$ & $239.52 \pm 0.45$ \\
15 & 210 & $14.87 \pm 0.10$ & $210.20 \pm 0.30$ \\
20 & 180 & $19.85 \pm 0.07$ & $179.79 \pm 0.22$ \\
25 & 150 & $25.02 \pm 0.08$ & $149.89 \pm 0.17$ \\
30 & 120 & $30.05 \pm 0.16$ & $119.85 \pm 0.14$ \\
35 & 90 & $35.04 \pm 0.20$ & $89.98 \pm 0.11$ \\
40 & 60 & $39.91 \pm 0.13$ & $59.94 \pm 0.10$ \\
45 & 30 & $44.99 \pm 0.05$ & $30.04 \pm 0.08$ \\
\hline \hline
\end{tabular}

\end{center}
\end{table}
\begin{table}
\caption{       \label{tbl4}
$xy$-vertex position uncertainty due to multiple Coulomb scattering by
track reconstruction procedure.
}
\begin{center}
Table shows the $xy$-vertex uncertainty ($\mu m$) obtained by a linear
track reconstruction with a 5 detector layers system, interspaced by
$dz$=1.5 cm, for some $z_1$ target to first detector layer distance (cm)
and some MUON momentum (MeV/c) values.\\
\bigskip
\begin{tabular}{||c|ccccccc||} \hline \hline
$z_1(cm)$ & 10 & 30 & 50 & 70 & 90 & 110 & 130 \\ \hline
$p(MeV/c)$ &&&&&&& \\ \cline{1-1}
500 & 203.414 & 600.261 & 997.307 & 1394.383 & 1791.468 & 2188.558 & 2585.650
\\
1000 & 116.415 & 336.970 & 557.795 & 778.662 & 999.541 & 1220.428 & 1441.317 \\
1500 & 86.285 & 245.645 & 405.317 & 565.036 & 724.772 & 884.515 & 1044.262 \\
2000 & 70.975 & 198.945 & 327.254 & 455.615 & 583.994 & 712.382 & 840.774 \\
2500 & 61.993 & 171.278 & 280.916 & 390.610 & 500.322 & 610.044 & 719.770 \\
3000 & 56.277 & 153.481 & 251.041 & 348.659 & 446.296 & 543.943 & 641.594 \\
3500 & 52.429 & 141.375 & 230.675 & 320.033 & 409.410 & 498.797 & 588.189 \\
4000 & 49.726 & 132.790 & 216.202 & 299.672 & 383.161 & 466.660 & 550.164 \\
4500 & 47.761 & 126.498 & 205.575 & 284.709 & 363.863 & 443.026 & 522.194 \\
5000 & 46.291 & 121.760 & 197.561 & 273.417 & 349.292 & 425.177 & 501.067 \\
5500 & 45.166 & 118.111 & 191.379 & 264.701 & 338.043 & 411.394 & 484.749 \\
6000 & 44.287 & 115.246 & 186.519 & 257.846 & 329.191 & 400.546 & 471.905 \\
6500 & 43.588 & 112.957 & 182.634 & 252.363 & 322.110 & 391.867 & 461.628 \\
7000 & 43.024 & 111.103 & 179.482 & 247.914 & 316.364 & 384.822 & 453.285 \\
7500 & 42.562 & 109.581 & 176.894 & 244.258 & 311.640 & 379.031 & 446.426 \\
\hline \hline
\end{tabular}

\end{center}
\end{table}
\vfill
\pagestyle{empty}
\newpage
\section*{\bf Figure captions}
\renewcommand{\labelenumi}{Fig. \arabic{enumi}}
\begin{enumerate}
\item   
     The track particle position  uncertainties  due  to  multiple
     Coulomb  scattering.  The  uncertainty  is   defined   as   the
     coordinate distribution width $\sigma_i$  on every detector layer $i$
     and results as the squares sum of all  preceding  layer  scattering
     contributions projected onto the $i$-th layer.
     $\sigma_{x_i}^2=\theta_0^2 [(z_i-z_1)^2+(z_i-z_2)^2+...+(z_i-z_{i-1})^2]$
\item   
     The plan projected x-coordinate point  distributions  of  the
     scattered 500~MeV/c MUONS, incident on detector layer No. 2, 3, 4,
     and 5 obtained by Monte-Carlo particle transport simulation.
\item   
Multiple Coulomb scattering error correlations.~~~~~~~~~~~~~~~~~~~~~~~~~~~~~~
\item   
     The $xy$-vertex position uncertainty (reconstructed vertex resolution
     deterioration) due to multiple Coulomb scattering as a function
     of MUON momentum and distance from  the  interaction  point  to
     first detector layer ($z_1$).
\end{enumerate}
\vfill

\end{document}